\documentclass[12pt]{iopart}
\usepackage{graphicx} % Include figure files

\usepackage {bm}% bold math

\begin{document}
\newcommand{\Par}[1]{\left (#1 \right )}
\newcommand{\om}{\omega}

\title{Mechanical feedback in the high-frequency limit}

\author{R.~El Boubsi, O.~Usmani, and Ya.~M.~Blanter}
 \address{Kavli Institute of Nanoscience, Delft University of
 Technology, Lorentzweg 1, 2628 CJ Delft, The Netherlands }

\date{\today}

\begin{abstract}
We investigate strong mechanical feedback for the
single-electron tunneling device coupled to an underdamped  harmonic
oscillator in the high-frequency case, when the mechanical energy
of the oscillator exceeds the tunnel rate, and for weak coupling. In
the strong feedback regime, the mechanical
oscillations oscillated by the telegraph signal from the SET in their
turn modify the electric current. In contrast to the earlier results
for the low frequencies, the current noise in not enhanced above the
Poisson value. 
\end{abstract}

\maketitle

\section{Introduction}

One of the key questions in the field of nanoelectromechanical systems
(NEMS) \cite{Cleland} is the effect of mechanical motion on the
electric properties of the systems. It is equally important for
undestanding of fundamental questions such as behavior of
non-equilibrium, dissipative, driven systems and for the prospects of
NEMS practical applications, such as switches, relays, and
actuators. Electron motion in both external ac electromagnetic fields
(see {\em e.g.} Ref. \cite{TienGordon}) and subject to time-dependent
noise \cite{CB1} have been extensively studied in the past. The new
feature brought by NEMS is that not only electrons move in the
field created by the mechanical vibrations (for brevity, phonons), but
also the phonons are created by the electron motion through the NEMS
device. The phonons are generally driven out of equlibrium thus
creating the feedback for the electron motion. 

In this Article, we concentrate on NEMS operating in the single
electron tunneling (SET) regime. These include experimental
realizations based on single molecules \cite{molecules}, semiconductor
beams \cite{Knobel,Weig,LaHaye}, and suspended carbon nanotubes
\cite{Sapmaz1,LeRoy,Sazonova,Witkamp,Bachtold}. In SET 
regime a NEMS can be represented as a SET device coupled to a
mechanical (harmonic) oscillator. The coupling is provided by a
force (typically of electrostatic origin, see {\em e.g.}
Ref. \cite{Sapmaz}) acting on the oscillator. The value of the force
depends on the charge state of the SET device, providing the
feedback. This feedback can be strong even if the electron-phonon
coupling is weak: Indeed, if one consideres an SET next to the Coulomb
blockade threshold, only two charge states, say, $n = 0$ and $n=1$,
with $n$ being the number of electrons at SET, are important. The
occupation of SET above the Coulomb blockade threshold fluctuates
between zero and one, providing a force which is a random telegraph
signal. The signal swings an underdamped oscillator to large
amplitudes, which in their turn affect the transport properties of the
SET device by modifying the tunnel matrix elements and the energy
difference between the two charge states. 

The transport properties of NEMS devices depend on a number of
parameters. One is the coupling strength $g$, defined as the
square of the ratio of the displacement of the oscillator center
under the action of the force at the amplitude of the zero-point
motion, $g = F^2/(\hbar M \omega^3)$, where $M$ and $\omega$ are
the mass and the frequency of the oscillator, and $F$ is the
difference of the forces acting on the oscillator in the charge
states $n=1$ and $n=0$. In is important that the Planck constant
is in the denominator, and thus strong coupling actually means
quantum regime. Other parameters determining the behavior are
given by ratios of various energy scales characterizing the NEMS
device. The most relevant one is the ratio between the typical
tunnel rate $\Gamma$ and the oscillator frequency $\omega$.

Surprisingly, the strong-coupling (quantum regime) is more extensively 
studied 
in the literature. Assuming the oscillator is underdamped, for
$g \geq 1$ and $\omega \gg \Gamma$, the behavior of the system is
dominated by Franck-Condon physics. The tunnel rates are modified due
to emission of phonons in the course of electron tunneling
\cite{CB1}. For the coupling to a single mode, this leads to the steps
in the current as the function of the applied bias voltage, the height
of the step is determined by the coupling constant
\cite{Flensberg,Oppen,Oppen1}. For $\omega \ll \Gamma$ the electron
level 
width becomes bigger than the distance between the steps, and
Franck-Condon structure disappears. However, in this regime the
motion of the oscillator is much slower than the electron tunneling,
and one can use Born-Oppenheimer approximation, considering electrons
in the quasi-stationary field potential provided by the oscillator. In
this situation, the oscillator may become bistable, and the electron
tunneling is dominated by switching events between the two states of
the oscillator \cite{Mitra,Mozyrsky,Mozyrsky1}. All these phenomena
are the 
manifestation of strong 
mechanical feedback. Similar effects have been studied in
Ref. \cite{Clerk} in the context of superconducting NEMS. 

It is less obvious that strong feedback is also possible at weak
coupling $g \ll 1$. This is the classical regime, where Boltzmann
equation serves as the starting point
\cite{Armour,Isacsson}. Ref. \cite{Usmani} studied the low-frequency
case
$\omega \ll \Gamma$. The oscillator motion in this case is described
by Fokker-Planck equation with effective diffusion and effective
damping (originating from the electron tunneling out of the SET
device), both determined by the energy dependence of the tunnel
rates. It turns out that the strong feedback regime is feasible, but
the behavior of the current strongly depends on the energy dependence
of the tunnel rates. For instance, in the two most commonly
investigated cases --- electron tunneling through a single level and
electron tunneling through a continuum of levels with the constant
density, like in a single electron transistor, the strong feedback
does not appear. Four distinct regimes have been identified: (i)
no phonons generated; (ii) the oscillations are
generated with the fixed finite amplitude; (iii) the oscillator is
bistable, one state has the oscillations with zero amplitude, and
another one the oscillations with a finite amplitude; (iv) the system
is bistable, with the two states representing the oscillations with
two different amplitudes. In the regimes (ii), (iii), and (iv) current
is strongly modified with respect to the case when the phonons are not
generated. A quantity even more sensitive to the oscillations is the
current noise. The natural measure of the current noise in SET devices
is Poisson value of the zero-frequency spectral density $S$, $S_P =
2eI$, $I$ being the average current \cite{NoiseReview}. It turns out
that the noise is strongly enhanced avove the Poisson value in the
regimes (ii), (iii), and (iv), and may even become super-Poissonian in
the regime (i), when the current is not renormalized. 

In this Article, we consider the only regime not addressed so far:
weak coupling $g \ll 1$ and high frequency $\omega \gg \Gamma$. We
show that it shares many features with the low-frequency classical
regime described above. Even though we have not been able to identify
any bistable regimes, we still find the two regimes of zero and finite
amplitude, the latter demonstrating strong mechanical
feedback. We find that the noise, in contrast to the
low-frequency case, is always sub-Poissonian.

\section{Boltzmann equation}

At weak coupling, the motion of the oscillator is classical, and the
behavior of the system is characterized by the joint distribution
function, $ P_n (x,v,t)$. Here, $n$ is the charge state of the SET. We
assume that the SET is biased close to the edge of one of the Coulomb
diamonds, so that only two charge states are important for transport,
for definiteness $n=0$ and $n=1$. Furthermore, $x$ and $v$ are the
coordinate and the velocity of the oscillator. The starting point of
our classical approach is the Boltzmann (master) equation for the
distribution function \cite{Armour,Usmani}, 
\begin{eqnarray}
\label{mastereq}
\frac{\partial P_n}{\partial t}&+&
\left \{  v \frac{\partial}{\partial x}+
  \frac{\partial}{\partial
v}\frac{\cal{F}}{M}\right \}P_n - \mbox{St}\ [P] =0;
\\
\cal{F}&=& - M \om^2 x - \frac{M \om v}{Q}+
F_n \; ; \label{force}\\
\mbox{St}\ [P] & =  &(2n-1) \left ( \Gamma^{+}(x) P_0
- \Gamma^{-}(x) P_1 \right ) \ , \label{rates2}
\end{eqnarray}
which holds for an arbitrary relation between $\omega$ and
$\Gamma$. Here, the total force $ \cal{F}$ acting on the oscillator 
is the sum of the elastic force, friction force, and
charge-dependent coupling force,
respective to the order of terms in Eq. (\ref{force}), $Q \gg 1$ is
the quality factor. We count the position of the oscillator from its
equilibrium position in the $n=0$ state. In this case, $ F_n =n F$ . 
The "collision integral" $\mbox{St}\ [P]$ in the right-hand side
represents single electron tunneling.
There are four tunnel rates,
$ \Gamma_{L,R}^{\pm}$ , where the subscripts $ L $ and $ R $ denote
tunneling through the left or right junction, and the superscripts $ + $
and $ - $ correspond to the tunneling to and from the island,
respectively; $ \Gamma^{\pm}= \Gamma_L^{\pm}+
\Gamma_R^{\pm}$. Each rate is a function of
the corresponding energy cost $ \Delta E_{L,R}^{\pm}$ associated with the
addition/removal of an electron to/from  the island in the state
$ n=0/1 $ via left or right junction
($ \Delta E_{L,R}^{+}= - \Delta E_{L,R}^{-}$ ). Two independent
energy differences are
determined by electrostatics and depend linearly on the
voltages. There is also a contribution to each energy from the shift
of the oscillator, $\Delta E_{L}^{+}=\Delta E_L^{+(0)}-Fx$, $\Delta
E_{R}^{-}=\Delta E_R^{-(0)}+Fx$, where $\Delta E^{0}$ are the
corresponding energy differences in the absence of mechanical motion. 

The condition $\omega \gg \Gamma_t \equiv \Gamma^+ + \Gamma^-$ means
that the motion of the 
oscillator is very slow compared with the typical time an electron
spends in the SET device. In this situation, the probabilities $P_0$
and $P_1$ average over the fast osicllator motion. In the leading
order, if we parameterize $x = \varepsilon \sin \varphi$, $v =
\omega \varepsilon \cos \varphi$, $\varepsilon = \sqrt{2E/M\omega^2}$,
these probabilities do not depend on
$\varphi$. Consequently, we expand the probabilities in the following
way,
\begin{displaymath}
P_n (x,v,t) \approx P_n^{a} (E,t) + \cos \varphi P_n^{b} (E,t) + \sin
\varphi P_n^{c} (E,t) \ ,
\end{displaymath} 
with $P_n^{b,c} \ll P_n^a$. Here and below we disregard the terms
proportional to $\sin m\varphi$, $\cos m\varphi$, with $m \ge 2$. This
procedure is similar to the transformation of Boltzmann equation into
the diffusion equation in the semi-classical theory of electron
transport in metals. 

We can obtain a closed set of equations for $P_n^{a,b,c}$ by
multiplying Eq. (\ref{mastereq}) with $1$, $\cos \varphi$ and $\sin
\varphi$ and subsequently averaging over the phase, throwing out $\sin
2\varphi$ and $\cos 2\varphi$ terms. It is important that the tunnel
rates only depend on the coordinate and not of velocity of the
oscillator, and thus are functions of $\sin \varphi$ and not $\cos
\varphi$. After some algebra, we obtain $P_{0,1}^b = P_0^c = 0$,
\begin{eqnarray*}
P_0^a (E,t) & = & \frac{\langle \Gamma^- \rangle}{\langle \Gamma_t
  \rangle} P(E,t), \ \ \ P_1^a (E,t) = \frac{\langle \Gamma^+
  \rangle}{\langle \Gamma_t \rangle} P(E,t),  \\
P_c^1 (E,t) & = & \frac{\langle \sin \varphi \Gamma^+ \rangle \langle
  \Gamma^- \rangle - \langle \sin \varphi \Gamma^- \rangle \langle
  \Gamma^+ \rangle}{\langle \Gamma_t \rangle \langle \sin^2
  \varphi \Gamma^- \rangle} P(E,t) \ , 
\end{eqnarray*}
where the angluar brackets denote averaging over the phase $\varphi$,
and the function $P(E)$ obeys the equation
\begin{eqnarray} \label{mastereq1}
& & \frac{\partial P}{\partial t} = {\cal L} P \equiv \sqrt{\frac{g\hbar
    \omega^3 E}{2}}\frac{\langle \Gamma^+ \rangle}{\langle \Gamma_t
    \rangle} \\
& \times & \left( -\frac{\partial P}{\partial E} + \sqrt{\frac{1}{2g\hbar
    \omega E}} \frac{\langle \sin \varphi \Gamma^+ \rangle \langle
  \Gamma^- \rangle - \langle \sin \varphi \Gamma^- \rangle \langle
  \Gamma^+ \rangle}{\langle \Gamma^+ \rangle \langle \sin^2
  \varphi \Gamma^- \rangle} P \right) \ , \nonumber
\end{eqnarray}
and we have disregared the mechanical damping $Q^{-1}$. 

\section{Distribution and current}

The stationary solution of Eq. (\ref{mastereq1}) is easily found as
\begin{eqnarray} \label{stationary}
& & P(E) = P(0)\exp \left( - \int_0^E \gamma(E') dE' \right)\ , \\
& & \gamma (E) = \sqrt{\frac{1}{2g\hbar
    \omega E}} \frac{\langle \sin \varphi \Gamma^+ \rangle \langle
  \Gamma^- \rangle - \langle \sin \varphi \Gamma^- \rangle \langle
  \Gamma^+ \rangle}{\langle \Gamma^+ \rangle \langle \sin^2
  \varphi \Gamma^- \rangle} \ . \nonumber
\end{eqnarray}

Note that, similarly to the low-frequency case $\omega \ll \Gamma_t$,
the distribution (\ref{stationary}) is very sharp. Indeed, the typical
energy $E$ is of the order of the applied voltage $eV$, and all
energies which occur in Eq. (\ref{stationary}) are in our classical
consideration much bigger that the quantum energy $\hbar \omega$ of
the oscillator. The function $\gamma$ can have positive as well as
negative sign. If it becomes negative at some values of energy
(amplification instead of the dissipation), there is a possibility
that the strong feedback regime emerges. 

\begin{figure}
\includegraphics{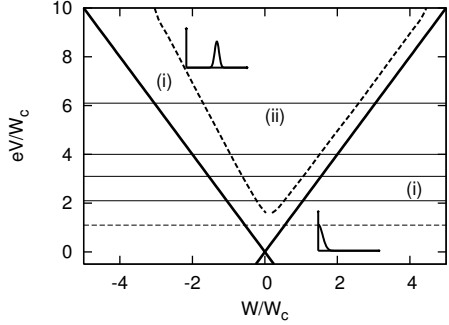}
\caption{\label{fitzones}
The stability regions in the gate-bias voltage plane.
Bold solid lines indicate the edge of the Coulomb diamonds.
Insets show the sketch of $P(E)$ in each region.
The  horizontal lines indicate bias voltages used for
current and noise scans in Figs.~2 and 3.
}
\end{figure}

To illustrate the energy dependence of the probability, we have chosen
exponential energy dependence used previously for low frequencies in
Ref. \cite{Usmani},
\begin{eqnarray}\label{rates1}
\Gamma_{L,R}^{+}& = & 2\Gamma_{L,R}^0 e^{-a_{L,R}\Delta E_{L,R}^{+}}
(1-f_F (-\Delta E_{L,R}^{+})) \ ; \nonumber \\
\Gamma_{L,R}^{-}& = & \Gamma_{L,R}^0 e^{a_{L,R}\Delta E_{L,R}^{-}} 
f_F (\Delta E_{L,R}^{-}) \ ,
\end{eqnarray}
the factor $2$ accounting for the spin degeneracy of the state $n=1$.
For concrete illustration, we choose  $ a_L = 0.3$ , $ a_R =
0.75$ , and $ \Gamma_{L}^0=\Gamma_R^0 $. In the figures, $W = eV_g$,
the gate $V_g$ and bias $V$ voltages are
measured in the units of $W_c$, the parameter associated with the
energy dependence of the tunneling rates, with the value smaller than
the charging energy.

Fig. 1 presents the regions in gate-bias voltage plane corresponding
to the two regimes. In the regime (i), the probability distribution
has a sharp peak around zero energy. Since the
energy represents the mechanical motion of the oscillator, 
it corresponds to the positive
damping at any energy and the absence of mechanical feedback ---
phonons are not generated by the electron tunneling events. In the
regime (ii), the distribution function sharply peaks around a finite
value of the oscillator energy: Mechanical oscillations are
generated. We were not able to detect the existence of
the bistable regimes, similar to (iii) and (iv) described in the
Introduction. The regime (ii) of phonon generation only emerges
outside the Coulomb diamonds. 

For (i), the distribution function can be approximated as 
\begin{equation} \label{zeropeak}
P(E) = \gamma (0) \exp (-\gamma (0) E) \ ,
\end{equation}
where we have normalized the solution. For (ii), we have the Gaussian
centered around the most probable value $E_m$,
\begin{equation} \label{onepeak}
P(E) = \frac{1}{\sqrt{2\pi}} \sqrt{\frac{\partial \gamma
(E_m)}{\partial E}} \exp 
\left( -\frac{\partial \gamma (E_m)}{\partial E} \frac{(E - E_m)^2}{2}
\right) \ . 
\end{equation}

The current is found as
\begin{equation} \label{curhighfreq}
I = \int I(E) P(E) dE, \ \ \ I(E) \equiv e \frac{\langle \Gamma_L^+
  \rangle \langle \Gamma_R^- \rangle - \langle \Gamma_L^-
  \rangle \langle \Gamma_R^+ \rangle }{\langle \Gamma_t \rangle} \ .
\end{equation}
Fig.~2 shows the results for the voltage dependence of the
current. The trace (a) is taken in the regime (i), and the current is
not modified by mechanical motion. The traces (b), (c) and (d) cross
the region (ii), and the current dependence in this regime deviates
from the one without mechanical motion. The deviations are stronger
for higher bias voltages; additional peaks in the current develop.

\begin{figure}
\includegraphics{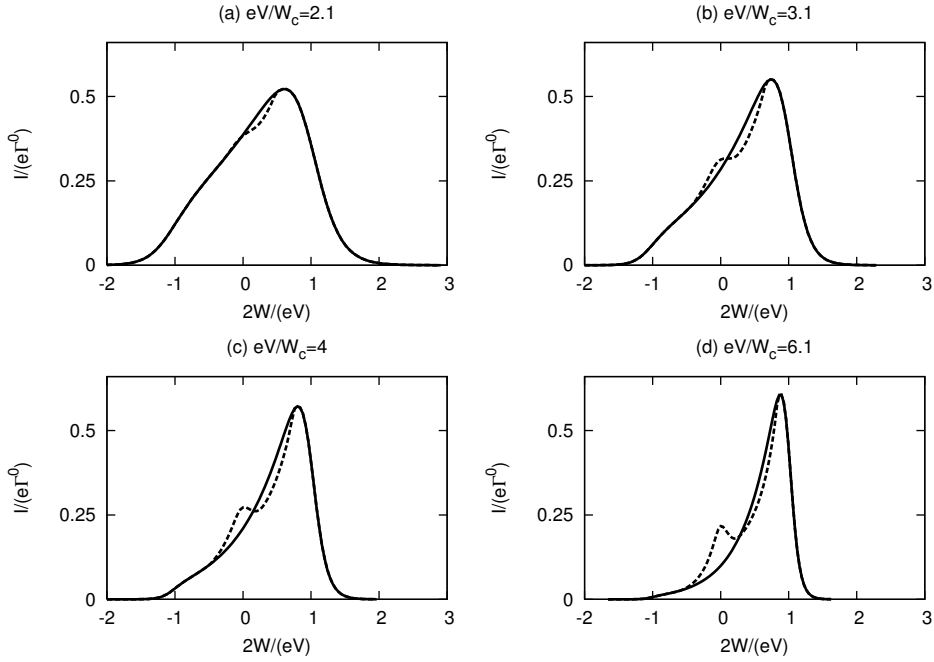}
\caption{\label{currents}
Current modification in strong feedback regime for different bias
voltages. 
The dashed (solid) lines give the current
modified  (unmodified) by mechanical motion. 
The modification is restricted to the regime
to region (ii) where the phonon generation of takes place.
}
\end{figure}

\section{Current noise}

The current noise spectral power at zero frequency is found from the
expression \cite{Usmani}
\begin{equation} \label{curnoise}
S = -4 \int_0^{\infty} \delta I (E) u(E) dE, \ \ \ \delta I (E) \equiv
I(E) - \int_0^{\infty} I(E) P(E) dE \ ,
\end{equation}
and $u$ solves the equation ${\cal L} u = \delta I (E) P(E)$. In
contrast to the low-frequency case, ${\cal L}$ is a first-order
differential 
operator. This fact simplifies the calculations and readily provides
analytical estimates for noise in the two regimes. We parameterize
$u(E) = v(E) P(E)$, and the equation for $v$ reads
\begin{displaymath}
-\sqrt{\frac{g\hbar \omega^3 E}{2}} \frac{\langle \Gamma^+
 \rangle}{\langle \Gamma_t \rangle} \frac{\partial v}{\partial E} =
 \delta I (E) \ .
\end{displaymath}
Solving it and substituting the result to the expression for noise, we
obtain 
\begin{equation} \label{noisegen}
S = 4 \sqrt{\frac{2}{g\hbar\omega^3}} \int_0^{\infty} \delta I (E)
P (E) \int_0^E \frac{\langle \Gamma_t \rangle}{\langle \Gamma^+
\rangle} \delta I (E') dE' \ .
\end{equation}
This expression can be evaluated with the use of the
approximations (\ref{zeropeak}), (\ref{onepeak}).

In the regime (i), we write $\delta I (E) \approx (g \hbar
\omega/2e^2) \partial^2 I/\partial V_g^2 (E - \gamma^{-1} (0))$,
where the second derivative is evaluated at $E = 0$. Substituting
this into the expression for noise, we obtain
\begin{equation} \label{noiseno1}
S = \sqrt{\frac{2\pi}{\omega}} (g \hbar \omega_0)^{3/2} \left. \left(
\frac{\partial^2 I}{e^2 \partial V_g^2} \right)^2 \right\vert_{E=0}
\frac{\langle \Gamma_t \rangle}{\langle \Gamma^+ \rangle}
(\gamma(0))^{-5/2} \ .
\end{equation}
The comparison with the Poisson value of noise $S_P = 2eI$ gives the
estimate $S/S_P \sim (\Gamma_t^2/\omega^2)(g\hbar\omega/eV_g)^{3/2}$. 
In the high-frequency regime $\Gamma_t \ll \omega$,
both factors are small, and thus in the
regime (i) the mechanically-induced noise is always sub-Poissonian.

In the regime (ii), expanding $\delta I = (\partial I (E_m)/\partial
E) (E - E_m)$, we obtain
\begin{equation} \label{noiseno2}
S = \frac{16}{3\omega} \sqrt{\frac{2 E_m}{g\hbar\omega}} \left. \left(
\frac{\partial I}{e\partial V_g} \right)\right\vert_{E = E_m}
\left. \left( 
\frac{\partial \gamma}{\partial E} \right)^{-1} \right\vert_{E = E_m}
\frac{\langle \Gamma_t \rangle}{\langle \Gamma^+ \rangle} \ .
\end{equation}
The estimate for the noise power is $S/S_P \sim
(\Gamma_t^2/\omega^2)(g\hbar\omega/eV_g)^{1/2}$. It is not surprising
that the noise is relatively higher than in the regime (i), where
there is no mechanical motion induced. However, the noise is still
below the Poisson value, which means that the behavior of the current
noise is dominated by the shot noise.

\begin{figure}
\includegraphics{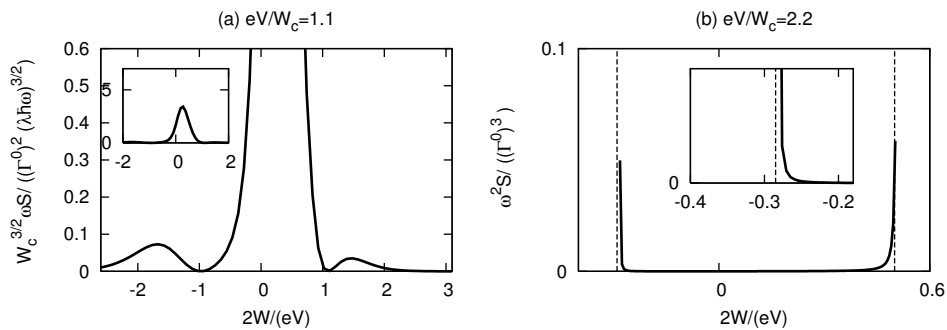}
\caption{\label{noise}
Mechanical contribution to current noise for different bias voltages.
(a) region (i); (b) region (ii).
}
\end{figure}

\section{Conclusions}

In this Article, we studied current and current noise for an SET
device coupled to an 
underdamped harmonic oscillator in the only regime not considered so
far: weak coupling $g \ll 1$ and high frequency $\omega \gg
\Gamma_t$. We find that, similarly to other regimes, coupling to
mechanical modes of the oscillator -- phonons -- excited by the
tunneling of electrons through the SET device, may have a strong
effect on the transport properties of the SET device. This is the
phenomenon of strong mechanical feedback. 

However, we also find that the feedback effects are the weakest of all
the regimes. It was very much expected that the effect of phonons on
electron transport is stronger for stronger coupling. But we also find
that for high frequencies the effects are less pronounced than for low
frequencies. Indeed, no bistability regimes have been discovered, the
strong feedback only manifests in oscillations with a fixed
amplitude. The mechanical contribution to the current noise is small
as compared with the shot noise contribution. It still can be
serparated from the white shot noise due to its frequency dependence,
however, the noise behavior is less spectacular than for low
oscillator frequencies, where it is sometimes exponentially enhanced
in comparison with the Poisson value. 

In this Article, we assume that the energy dependence of the tunnel
rates is not too strong. All our results are expressed in terms of the
tunnel rates averaged over the period of mechanical oscillations. One
can imagine an opposite situation --- strongly energy-dependent rates
(but at each energy still below the frequency of the
oscillator). Since the oscillator moves quickly on the scale of the
typical time the electron spends in one of the SET charge states, the
instant tunnel rate performs fast oscillations with a big
amplitude. In this situation, it is the easiest for an electron to
tunnel when the tunnel rate is the highest, which typically would
correspond to the maximum displacement of the oscillator. In this {\em
synchronization} regime, the electron jumps are synchronized with the
oscillator period. The condition for the appearance of the
syncrtonization regime is $(d\Gamma/dE) \delta E \gg \Gamma$, where
$\delta E \sim (eV)^2/(g\hbar\omega)$ is the shift of the mechanical
energy if the oscillator is displaced between the two extreme
positions. Presumably, in the syncronization regime the transport
properties are determined by the maximum (rather than the averaged)
value of the tunnel rate over the oscillation period. Detailed
analysis of the sincronization regime lies outside the scope of this
Article, but we expect that the current is enhanced as compared with
the ``regular'' strong-feedback regime considered above, whereas the
current noise is suppressed since the electron stream becomes more
regular.

This work was supported by the Netherlands
Foundation for Fundamental Research on Matter (FOM).

\end{document}